\documentclass[conference]{IEEEtran}
\IEEEoverridecommandlockouts
\usepackage{cite}
\usepackage{amsmath,amssymb,amsfonts}
\usepackage{algorithmic}
\usepackage{graphicx}
\usepackage{textcomp}
\usepackage{xcolor}
\usepackage{comment}
\usepackage[margin=1.035in]{geometry}

\def\BibTeX{{\rm B\kern-.05em{\sc i\kern-.025em b}\kern-.08em
    T\kern-.1667em\lower.7ex\hbox{E}\kern-.125emX}}
\begin{document}

\title{\fontsize{22}{20}\selectfont

SWI-FEED: Smart Water IoT Framework for Evaluation of Energy and Data in Massive Scenarios


{
\footnotesize 
\textsuperscript{}}
\thanks{This work was supported by the European Union under the 
Italian National Recovery and Resilience Plan 
NRRP of NextGenerationEU, partnership on “Telecommunications of the Future” (PE00000001 - program “RESTART”), specifically in the WITS and SPRINT focused projects.}
}

\author{

\IEEEauthorblockN{ \small Antonino Pagano, Domenico Garlisi, Fabrizio Giuliano}
\IEEEauthorblockA{\textit{ \small University of Palermo (Palermo, Italy)}\\
\textit{ \small CNIT (Parma, Italy)}\\
{ \small name.surname}@unipa.it}
\and
\IEEEauthorblockN{\small Tiziana Cattai,  Francesca Cuomo}
\IEEEauthorblockA{\textit{ \small University of Rome "La Sapienza" (Rome, Italy)}\\
\textit{\small CNIT (Parma, Italy)}\\
{\small name.surname}@uniroma1.it}
}

\maketitle

\begin{abstract}
\fontsize{9}{9}\selectfont
This paper presents a comprehensive framework designed to facilitate the widespread deployment of the Internet of Things (IoT) for enhanced monitoring and optimization of Water Distribution Systems (WDSs). The framework aims to investigate the utilization of massive IoT in monitoring and optimizing WDSs, with a particular focus on leakage detection, energy consumption and wireless network performance assessment in real-world water networks. 
The framework integrates simulation environments at both the application level (using EPANET) and the radio level (using NS-3) within the LoRaWAN network. The paper culminates with a practical use case, alongside evaluation results concerning power consumption in a large-scale LoRaWAN network and strategies for optimal gateway positioning. 
\end{abstract}


\section{Introduction}
%
Water scarcity is a global threat, emphasizing the need for effective water management. Researchers are working to optimize water resource utilization due to increasing demand and climate change impacts. Enhancing the monitoring and the management of Water Distribution Systems (WDSs) can reduce resource wastage and the Low Power Wide Area Network (LPWAN) technologies represent a promising technology to deploy applications for monitoring and controlling smart systems on a large scale, leading to the development of Smart WDSs (SWDSs) \cite{velayudhan2022WITS4}.
However, implementing SWDSs requires two-level integration: a robust, secure communication infrastructure, and advanced applications for data analysis, to the aim to predict anomalies or resource wastage. This paper introduces SWI-FEED (Smart Water IoT-Framework for Evaluation of Energy and Data in Massive scenarios), an innovative holistic framework for smart water management assessment that addresses the integration challenges. The framework considers both the high-level and low-level network properties of a deployment, as well as its topological aspects.
Current literature uses both water and wireless simulators but not in an integrated way. To the best of our knowledge, there is no existing framework in the current literature that merges these two levels of integration SWDSs.

The contribution of SWI-FEED in the field of SWDSs is multifaceted and includes, but is not limited to, the following objectives: i) Establishing the Topology of the SWDS; ii) Assembling a Dataset; iii) Selecting Features, Representing Graph Topology, and Reducing Data Dimensions; iv) Testing Energy Reduction and Leakage Detection Algorithms; v) Evaluating Approaches in Large-Scale SWDS Scenarios.
The following sections will delve into the details of the proposed framework. Sec. \ref{sec:system-architecture} presents the system architecture. Sec. \ref{sec:methodology} describes the methodology employed in the framework. Finally, an example of the framework's application is presented in Sec. \ref{sec:usecase} together with the paper conclusion.

\section{System Architecture and Related work}
\label{sec:system-architecture}
%
\begin{figure}[!b]
    \centering
    \includegraphics[width=0.95\linewidth]{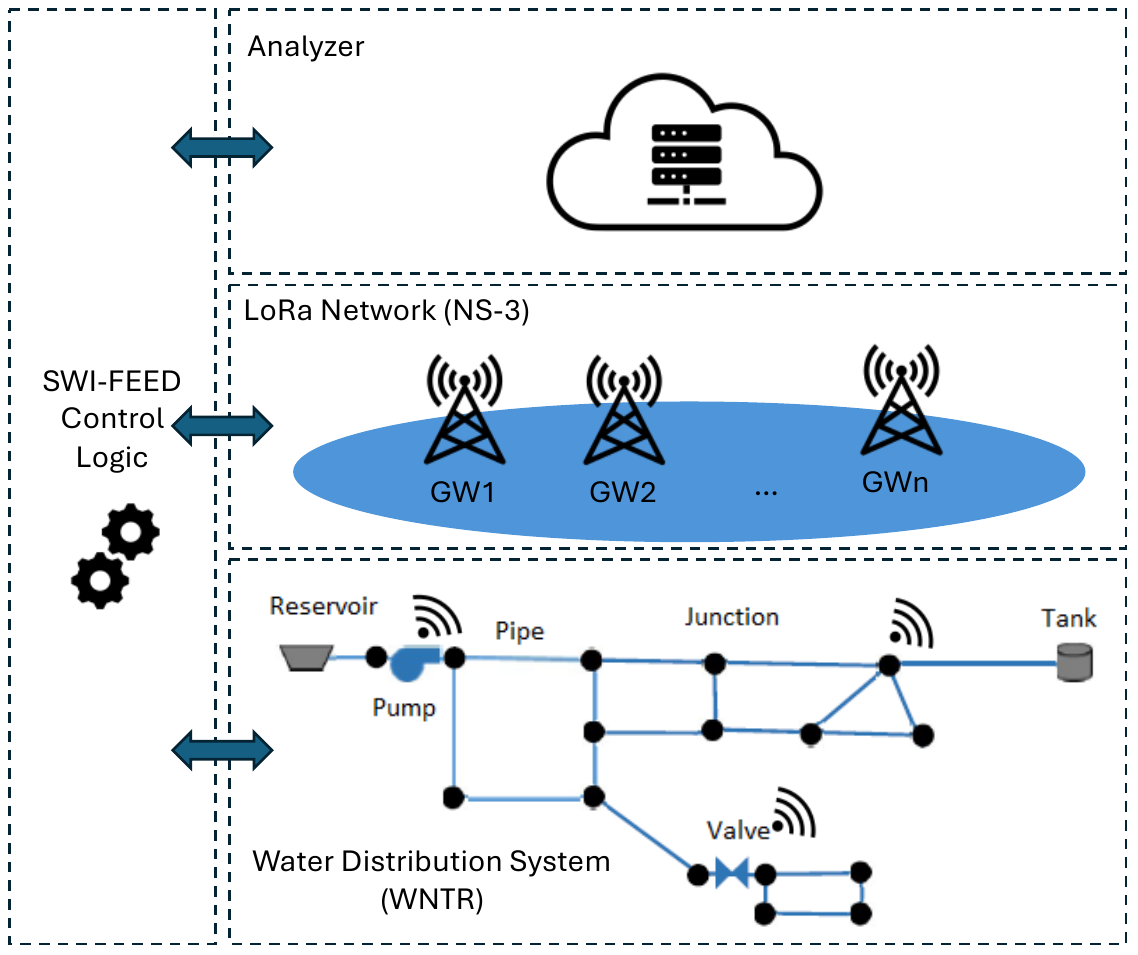}
     \caption{SWI-FEED Architecture}
    \label{fig:architecture}
\end{figure}
In this section, we outline the structure of our evaluation framework, detailing the interconnected modules and the relationships between them. The proposed architecture, depicted in Fig. \ref{fig:architecture}, begins from the bottom, by defining the network topology and the high-level data application using the EPANET core \cite{rossman_computer_1999}, a tool for modeling and analytical evaluation of water supply systems. It stimulates the hydraulics and water quality behavior of piping networks. EPANET's capabilities extend to the evaluation of pressure, flow rates, and other critical parameters. For our scope, we integrate in SWI-FEED the Water Network Tool for Resilience (WNTR) \cite{KLISE2017420}, a Python package to interface EPANET.

We utilize EPANET to generate a dataset that functions as an input for NS-3 \cite{ns3}, as well as establish the groundwork for the algorithms used within our framework, which will be detailed in the following section. This is depicted in the central and upper sections of Fig. \ref{fig:architecture}, respectively.
In the context of IoT networks, among various simulation tools, the open-source NS-3 network simulator is identified as the most suitable option \cite{lorasurvey}. It accurately models the behavior of radio frequency signals, transmission, propagation, interference, and other relevant parameters. NS3 enables the analysis of network performance and can include energy evaluation.
Moreover, among the various LPWAN technologies, the Long Range Wide Area Network (LoRaWAN) \cite{Alliance13} emerges as the most suitable candidate, owing to its advantages such as low power consumption and extensive coverage.


LoRaWAN networks are deployed in a star-of-stars topology, sensor nodes send data to gateways (GWs), which then forward it to a central server. Additionally, the Spreading Factor (SF) parameter, in the range SF7-SF12 determines the transmission speed, coverage, and energy consumption. Lower values provide reduced coverage and lower energy consumption for sending a single packet but offer faster transmission speeds, while higher values increase coverage at the expense of higher energy consumption and reduced transmission speed. Furthermore, Adaptive Data Rate (ADR) is a feature that automatically optimizes the allocation of SFs based on signal conditions

Moreover, Fig. \ref{fig:architecture} shows the SWI-FEED control logic, it represents the engine to control the assessment process. It provides an easy structure to define the target performance and multiple evaluation scenarios, thus allowing the framework to automatically run experiments and adjust interested parameters. By iteratively exploring different combinations of parameters, SWI-FEED identifies the most effective configuration to achieve the desired KPIs. Indeed, integrating simulators from different application domains to assess SWDSs is a challenging task. While simulation systems are commonly used for studying communication networks, there are only a few studies that have integrated EPANET and NS-3 specifically for SWDS assessment \cite{garlisi2022WITS3}. These studies focus on IoT systems with sensors in WDSs but lack optimized integration of simulators from different domains.


\section{Methodology}
\label{sec:methodology}
\begin{figure}[!h]
    \centering
    \includegraphics[width=1\linewidth]{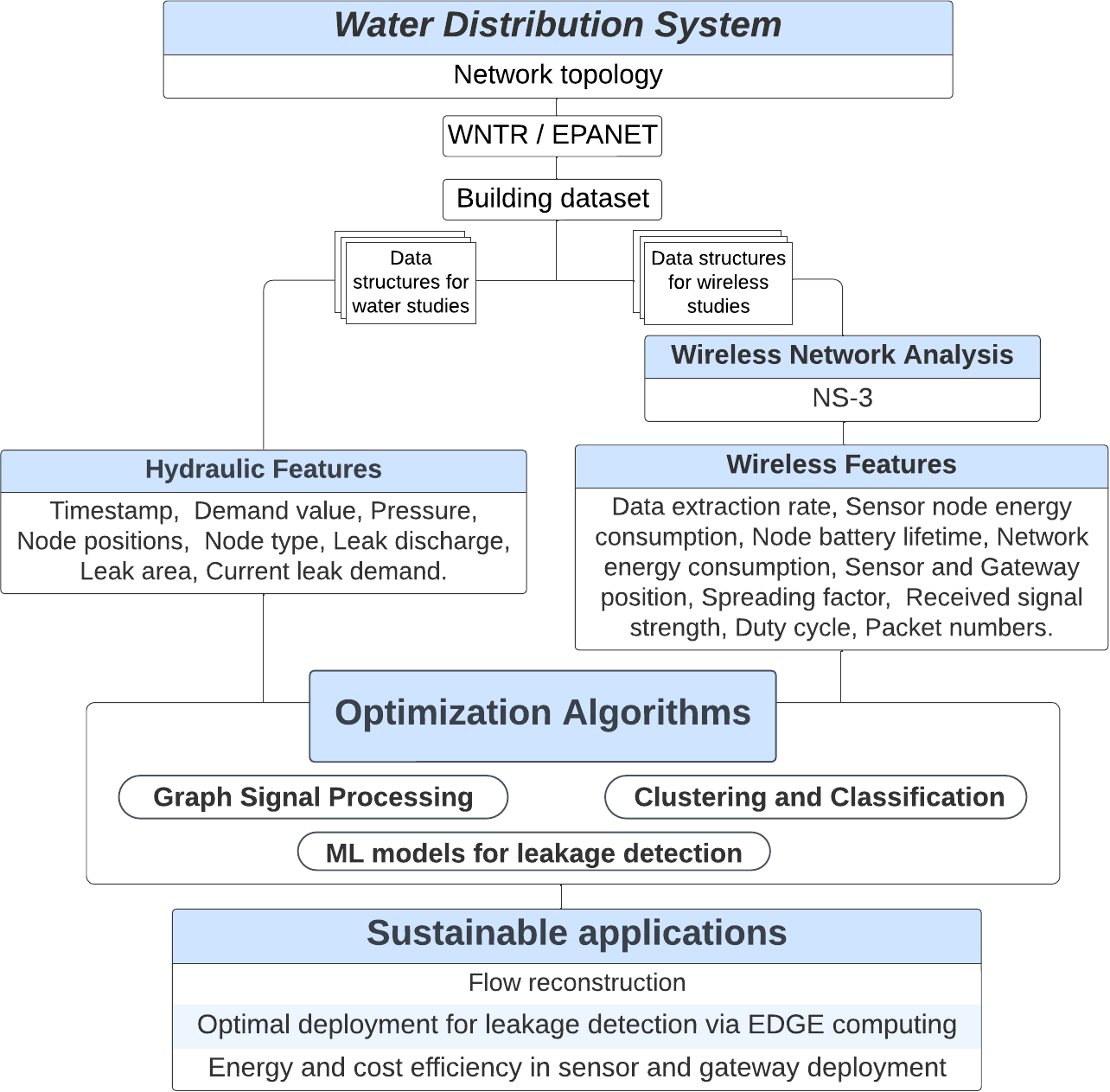}
    \vspace{-20pt}
    \caption{Framework for energy and data performance evaluation in Massive-IoT within SWDSs.}
    \label{fig:framework}
\end{figure}

\vspace{-6pt}
In this section, we provide a comprehensive description of the methodology applied from the proposed framework. The complete workflow, illustrated in Fig. \ref{fig:framework}, enables a holistic analysis of SWDSs by incorporating three main components: i) WDS network infrastructure and topology; ii) features analyzed; and iii) optimization algorithms.

\subsubsection{WDS network topology and Features Analyzed}
Moving from the top portion of the figure, the initial step in the methodology involves identifying the architectural model of the WDS. Subsequently, a dataset is created based on the network's topology. This task is accomplished using the WNTR/EPANET tools, which enable the analysis and simulation of the WDS. At this stage, the implemented SWI-FEED control logic comes into play, facilitating the automation of experiments and streamlining the integration into larger workflows. This control logic also facilitates the generation of datasets, which encapsulate the physical values of the WDS. If specified in the configuration, these datasets can be exported in CSV format for external use. The dataset comprises two types of data. The first type contains the structure for the application-level study, which includes the features listed in the left block of the figure, referred to as "Hydraulic Features". Furthermore, a subset of this dataset, containing information about node positions and sampling times, is utilized as input for the wireless network analysis block. This block performs an analysis of various parameters related to the IoT network, as listed in the right branch of the figure and referred to as "Wireless Features" 
Both, these data provide insights into the hydraulic and IoT network behavior of the SWDS and serve as a foundation for the application of the optimization algorithms.

\subsubsection{Optimization Algorithms and Sustainable applications}
In the second phase of the proposed methodology, various optimization algorithms are applied to process the hydraulic and IoT features obtained from the SWDS. 
The integration of these algorithms ensures sustainable network optimization, addressing both water management and wireless communication perspectives. For instance, ML algorithms have been employed in distributed analysis for leak detection in SWDS with EDGE computing \cite{garlisi2022WITS3}. By clustering nodes based on shared features, it becomes feasible to develop models capable of predicting behavior within specific subnetworks or clusters \cite{quinones2019novel}. These approaches allow for more accurate predictions and targeted interventions within localized areas of the network. Additionally, the utilization of classification and clustering algorithms facilitates the deployment of ML models tailored for leak prediction within clusters, further enhancing prediction accuracy \cite{chen2021iterative}.

\section{Use Case and Conclusion}
 \label{sec:usecase}

To demonstrate our methodology, we present a case study involving a massive IoT network supporting a WDS. This case study showcases the systematic approach for assessing and optimizing the performance of the SWDS, ensuring efficient water distribution and network operation. The specific focus of this case study is on evaluating strategies for LoRaWAN GWs deployment in alignment with the WDS hydraulic flow. 
The goal of the proposed Degree Centrality Deploy is to utilize the data and metrics produced in the preceding stages of the methodology to discern the relationship between centrality and hydraulic flow. This analysis will aid in determining the optimal locations for GWs.

In this context, we resort to the network theory, where a graph $\mathcal{G}$ is a composed by N nodes (or vertices) and links (or edges). Here, pipes can be represented as graph links while junctions as nodes. The adjacency matrix $A$ contains the interactions between nodes modeled in ways that depend to its definition. In this case, we consider the undirected adjacency matrix, where the generic element $a_{ij}=1$ if it exists an edge between node $i$ and $j$, otherwise $a_{ij}=0$.
With the aim to quantify the importance of the nodes in the network we consider the degree centrality \cite{freeman2002centrality}, defined for the i-node as $c_i = \frac{d_i}{N-1}$, where $d_i=\sum_{j=1}^{N} a_{ij}$ is the node degree of the i-node, that measure the number of connections containing the node i; $N$ is the total number of nodes of $\mathcal{G}$ and consequently $N-1$ corresponds to the maximum of degree of a N-nodes network. Node centrality measures are applied to determine the importance of each node within the WDS. We leverage the centrality associated with each node and the hydraulic flow to identify the optimal position for the network GWs.

Our use case focuses on a large network, which consists of 4419 nodes, 3 reservoirs, and 5066 pipes presented as topology and dataset at this link\footnote{https://github.com/WITS-Restart/WDN-IoT-Dataset-Workbench}.

The \mbox{SWI-FEED} examines the performance of the LoRaWAN network used to collect stream meter measures within this network placed at each node, particularly focusing on network energy consumption. Each device transmitted data at an average interval of 5 minutes and the battery level of each device was observed over a 24-hour experimental period.
\begin{table}[t!]
\centering
 \caption{ Comparison of daily network energy consumption between the two different GW deployments}\label{tab:comparison}
\resizebox{0.9\columnwidth}{!}{%
\begin{tabular}{|c|c|c|}
\hline
\textbf{\begin{tabular}[c]{@{}c@{}}GWs \\ Number\end{tabular}} & \textbf{\begin{tabular}[c]{@{}c@{}}Energy Consumption [J] \\ Regular Grid Deploy\end{tabular}}  & \textbf{\begin{tabular}[c]{@{}c@{}}Energy Consumption [J] \\ Degree Centrality Deploy\end{tabular}} \\ \hline
77 & 189574 & 147500 \\ \hline
96 & 161589 & 131959 \\ \hline
117 & 116585 & 98625 \\ \hline
140 & 96648 & 82626 \\ \hline
165 & 66621 & 60443 \\ \hline
\end{tabular}
}
\end{table}\\
Tab. \ref{tab:comparison} provides a comparison of daily network energy consumption between two different gateway distributions: Regular Grid Deploy and Degree Centrality Deploy. It can be observed that as the number of GWs increases, there is a tendency to decrease energy consumption for both deployment methods. This is because as the density of GWs increases, the ADR assigns lower SF values to the end devices. However, it is evident that the proposed method consistently exhibits lower energy consumption than Regular Grid Deploy in all configurations. This comparison sheds light on the potential energy efficiency benefits of using degree centrality for GWs distribution strategies in SWDSs.

At the final stage, this paper aims to provide a comprehensive and effective framework for assessing and improving SWDSs, together with a methodology that introduces new parameters and metrics that help performance evaluation, especially in massive IoT scenarios, thereby making it easy to compare different models. Finally, we have shown a use case scenario.

\bibliographystyle{IEEEtran}
\bibliography{biblio}
\end{document}